\documentstyle[aps,prb,epsf,amsmath,multicol]{revtex} 
\begin{document}
\draft
\title{Selforganized 3--band structure of the doped fermionic Ising spin
glass}
\author{H. Feldmann and R. Oppermann}
\address{Institut f\"ur Theoretische Physik, Universit\"at W\"urzburg,
97074 W\"urzburg, F.R.Germany}
\date{\today}
\maketitle

\newcommand{\qs}{\tilde{q}}
\newcommand{\chib}{\overline{\chi}}
\newcommand{\calC}{{\cal{C}}}
\newcommand{\erf}{{\rm erf}}

\begin{abstract}

The fermionic Ising spin glass is analyzed for arbitrary filling and
for all temperatures. A selforganized 3--band structure of the model is
obtained in the magnetically ordered phase. 
Deviation from half filling generates a central nonmagnetic band, which
becomes sharply separated at $T=0$ by (pseudo)gaps from upper and lower
magnetic bands.
Replica symmetry breaking effects are derived for several observables
and correlations.
They determine the shape of the 3-band DoS, and, for
given chemical potential, influence the fermion filling strongly in the
low temperature regime. 

\end{abstract}


\section{Introduction}
Magnetic correlations with frustration in insulating fermionic systems are
a key topic of modern condensed matter theory. In disordered systems with
frustrated random spin--spin interactions the description is found to be 
highly non perturbative,
comparable for example with problems encountered in disorder--free Hubbard
models. Limits, where mean field theories become exact and probably include
essential physics, exist, but their exact solutions are hard to obtain. 
These complications are fortunately also linked to the richness of
physical phenomena, already incorporated in mean field equations. 
The important experimental tool of doping often leads to quantum
phase transitions of various kinds and thus further enhances the structure
of phase diagrams. It is the purpose of the present paper to contribute to 
the understanding of the low temperature glassy phase,
which reacts strongly to doping before its random magnetic order is
destroyed in favor of a paramagnetic phase.\\   
Systems with magnetic interaction often realize saturated magnetic
order in the ground state at zero temperature, even if the interaction is
frustrated and random such that magnetic order consists of randomly
oriented frozen--in magnetic moments. 
It is well known that spin glass order requires a
mean field theory with more than one order parameter or even an order
parameter function \cite{parisi80b,parisi80c,binder86a,fischer91a} $q(x)$ 
with Parisi parameter $x$, in terms of which complete random order can be
identified: 
the support of the subset of order parameters, which do not reach their
maximum allowed value at $T=0$, shrinks to zero and the Edwards Anderson
order parameter reaches its maximum.
In case of interacting fermionic spins, the magnetic saturation can
be increasingly depressed by doping until a breakdown of magnetic
order occurs at sufficiently large deviation from half filling.
The chemical potential $\mu$ as a competing field, which breaks
particle--hole symmetry, and the fermion concentration $\nu(\mu)$ are
hence important parameters. 
A famous example for doping effects on magnetism is the
breakdown of antiferromagnetic order in Hubbard type models
\cite{georges96a}.
The doped fermionic Ising spin glass(${\it ISG_f}$), the properties of which
are at the center of the present paper, contains a couple of close
relationships with the Hubbard model.
Despite very different techniques employed in the two cases, e.g. the
replica formalism for the disordered model does not show up in the
clean Hubbard model, comparable phenomena in the band structures
appear and will be discussed.\\
In preceding papers, the density of states of the ${\it ISG_f}$ was analyzed in detail for half--filling
\cite{oppermann98c}. 
A pseudogap was obtained in a solution with infinitely many steps of
replica symmetry breaking (RSB). Hence the hierarchy of infinitely many
steps of breaking a discrete symmetry resembled the effect of breaking a
continuous symmetry, which commonly leads to soft modes. In the present
case, soft single particle excitation energies were created, irrespective 
of two particle soft modes, for which the single particle DoS appears only as
a weight.\\
Replica symmetry breaking introduces a sort of statistical fluctuation
effects in fermionic spin glasses, which determine the band shape. 
In particular the softening of gap energies occurs like a
onedimensional quantum critical phenomenon.\\
This single dimension can be viewed as the replica dimension, while a
similar role is played by the time in the dynamic mean field theory of the
infinite-dimensional Hubbard model. It was however also found for the
$ISG_f$, that the finer structures induced by symmetry breaking in
replica space were directly felt in the quantum dynamic behaviour of the
subclass of fermionic correlations.\\
The description of doping and of general arbitrary filling, which affects
size, position, and splitting of gaps, has been the goal of the
present work. The results of this paper are obtained by entangled (rather
than parallel) numerical and analytical analysis. This
`numerico-analytical' study is based on a replicated field theoretical
treatment of the random Ising interaction problem.\\
Doping and arbitrary filling enforces occupation of nonmagnetic states at
$T=0$, even where magnetic order is strongly preferred. 
The present paper involves and revisits earlier results for the
tricritical phase diagram \cite{oppermann96a} away from half--filling.\\
The low--temperature limit of the non--half filled model and the way the 
phase transition into the paramagnetic phase takes place, remained 
an open problem up to now.
This was related to several problems: first, information about
the full replica--broken solution (removing a negative
Almeida--Thouless (AT) eigenvalue) was only available at half--filling, and
secondly another AT eigenvalue turns complex due to the replica limit
and the question of stability is raised again.\\  
In this article, however, we suppose that standard replica symmetry breaking
is sufficient to describe the ordered phase. (see a more
detailed discussion of the stability properties in Refs.
\onlinecite{oppermann99a,oppermann99b}).\\  
After the presentation of the model and a short outline of the
calculation, we present first the results obtained within the 
replica--symmetric approximation.
The three band structure in the low--temperature regime is already
obtained in this lowest order approximation within a certain range of
chemical potentials. The formulas should be easily
understandable and helpful for any reader, who wants to limit his
interest to the selforganized appearence of the nonmagnetic third band; 
considerably increased efforts by one step RSB lead to a refined picture
of this 3-band structure. \\ 
We discuss the way these three bands emerge below the freezing
temperature and how they can be understood in terms of magnetic
and nonmagnetic contributions. It is well known that replica
symmetry breaking occurs and, reemphasizing that it has a profound effect
on the low temperature properties of the system, we show how it is manifested 
together with broken particle hole symmetry. For this purpose 
we present in detail calculations and results in one--step replica symmetry
breaking (1RSB). Fortunately, this solution already allows a sound
estimation of the exact one. 
Some key features suggested by this one--step RSB solution are also
confirmed by exact results for infinite RSB: among those, the
replacement of the hard gap by a pseudo gap and the onset of the three--band
splitting already at arbitrary small deviations from half--filling are
most important.\\
\section{The doped fermionic Ising spin glass}
Since in several preceding publications the model has been explained in
many respects, we wish to limit the present discussion to the specific aim
of this paper, hence the way doping interferes in the interplay of charge-
and spin-correlations. \\
The grandcanonical Hamilton operator
\begin{equation}
{\cal{K}}=\sum_{i<j}J_{ij}\sigma_i^z\sigma_j^z-\mu \sum_i n_{i\sigma}
\end{equation}
describes instantaneous interactions of fermionic spins. Stripping off the 
factor $\hbar/2$ the spin operators $\sigma^z$ are given by the fermion
particle number operators by $\sigma^z\equiv
n_{\uparrow}-n_{\downarrow}$. The distribution of random interaction
couplings
\begin{equation}
{\cal{P}}(J_{ij})=\frac{1}{\sqrt{2\pi}J} e^{-J_{ij}^2/(2 J^2)}
\end{equation}
generates magnetic correlations independent of time and of infinite
range in real space. 

\section{Magnetic and Nonmagnetic Bands}
\subsection{Replica Symmetric Results}
Despite its instability against replica symmetry breaking (RSB) the
replica symmetric solution is nontrivial and must be understood in
detail, since it forms the basis for the improved solution presented
below. Moreover, in a relatively simple description, this approximation
contains features of the selforganized transition from single-band
structure above freezing temperature to either two--bands for half filling
or three-band structure, which occur below freezing of the magnetic
moments. The quality of this approximation decreases in the low
temperature regime, as the next section will show. The role of symmetry
breaking amounts to a softening of the gaps and to a removal of sharp
drop-offs of the density of states at low temperatures, as demonstrated
earlier for the case of half-filling \cite{oppermann98c}. 
At half-filling a pseudogap results between upper and lower magnetic band. 
A special line in the phase diagram winding 
through the ordered phase, connecting tricritical point(s) and the zero  
temperature gap edges of the magnetic bands \cite{oppermann99a}, 
wraps a regime where the spectral weight of the central nonmagnetic
band does not fully move into the magnetic bands as $T\rightarrow0$. \\
In this section we first present selfconsistent numerical solutions for   
the temperature range from the filling-dependent freezing temperature down
to very low $T$ of order $10^{-4}J$. The numerical evaluation of the   
density of states employs numerical solutions of the selfconsistent
equations for spin glass order parameter $q$ and linear susceptibility   
$\chi$ as a function of temperature and parametrized by the chemical   
potential $\mu$. The fermion concentration is also calculated as a
function of $\mu$.
This analysis is supplemented by an exact $T=0$ calculation.
\subsubsection{The free energy and resulting selfconsistent equations at
finite temperatures}
The following results are derived from the effective Lagrangian and from
a generating functional for general fermionic correlation functions of
the model. Analytical solutions are found in the sense that the number of
nested integrations becomes minimized before, in a final step, the
observable are evaluated numerically. In order to facilitate the
$T\rightarrow0$ limit, it is useful to change variables by introducing the
linear susceptibility $\chi=\beta J (\qs-q)$. Since $\qs-q$ decays
linearly with $T$ in the $T\rightarrow0$ limit, the finite linear
susceptibility is a helpful quantity which reduces the degree of
$\beta$-divergences (which need to be compensated). For this reason $f$ is
expressed rather in terms of $q$ and $\chi$ than in terms of $q$ and $\qs$.\\ 
The free energy (density) $f$ is obtained in the replica symmetric
approximation as
\begin{align}
f&=\frac{1}{4}J\chi(\frac{\chi}{\beta J} + 2q - 2) - T \ln 2 - \mu
- T \int_z^G \ln \calC \quad \mbox{with}\\
\calC &=\cosh (\beta J \sqrt{q} z) + 
\cosh (\beta \mu)e^{-\frac{1}{2}\beta J\chi}
\label{finitetempF}
\end{align}
The Gaussian integral, which appears frequently throughout the paper, has 
been reduced to the short form defined by
\begin{equation}
\int_z^G \phi(z) \equiv\frac{1}{\sqrt{2\pi}}\int_{-\infty}^{\infty}dz
e^{-\frac12 z^2} \phi(z)
\end{equation}
Extremalization of expression (\ref{finitetempF})  with
respect to $q$, $\chi$, and $\mu$ yields the coupled selfconsistent
equations
\begin{equation}
0=\partial_q f=\partial_{\chi}f=\nu+\partial_{\mu}f
\end{equation}
which we solved numerically. 
The results are finally used in the band structure calculation. 

\subsubsection{Energy and selfconsistent equations in the $T=0$ limit}

The zero temperature limit deserves separate attention for several
reasons. As can be seen from the thermal free energy, cancellation of
divergences in the $\beta\rightarrow\infty$ limit trouble the numerical
work at low temperatures but nondivergent analytical results at $T=0$
provide a control.\\
For $T=0$, an analytical approximation was also obtained, applying an
expansion in powers of $\mu-\chi/2$. Good agreement was found almost up to
the magnetic breakdown.\\
As usual the equations are simplified in the $T=0$ limit and were
obtained by a variant \cite{daCosta94a} of the Sommerfeld method. For
the density of states the steepest descent method is used in addition.\\
We obtain for $\mu <\frac{1}{2}J\chi$:
\begin{equation}
E\equiv 
f(T=0)=\frac{1}{2}J\hspace{.1cm}\chi(q-1)-\mu-\frac{2}{\sqrt{\pi}}J\sqrt{q} 
\end{equation}
while for $\mu >\frac{1}{2}J\chi$, the $T=0$ energy becomes
\begin{equation}
E=\frac{1}{2}J\chi(q-1)-\mu-(\mu-\frac{1}{2}J\chi)\hspace{.1cm}\erf
\Bigl(\frac{1}{\sqrt{2q}}(\frac{\mu}{J}-\frac{\chi}{2})\Bigr)
-J\sqrt{\frac{2 q}{\pi}} e^{-\frac{1}{2q}(\frac{\mu}{J}
-\frac{\chi}{2})^2}
\end{equation}
Extremalization of these energies yields $T=0$ self--consistency
equations which couple the magnetic correlations $q$ and $\chi$
with the filling factor as a charge average.\\
For $\frac{\mu}{J} > \frac{\chi}{2}$ we derive the following relations
between zero temperature parameters
\begin{align}
q=&1-\erf\Bigl(\frac{1}{\sqrt{2
q}}(\frac{\mu}{J}-\frac{\chi}{2})\Bigr),
\quad
\chi=\frac{2}{\sqrt{2\pi q}}e^{-\frac{1}{2 q}(\frac{\mu}{J}
-\frac{\chi}{2})^2}\\
\nu=& 1 + \erf
\Bigl(\frac{1}{\sqrt{2 q}}(\frac{\mu}{J}-\frac{\chi}{2})\Bigr),\quad
\quad q = \qs = 2-\nu \quad .
\label{zeroTempeq1}
\end{align}
One may derive the $T=0$ solutions as a function of either $\nu$ or
$\mu$. The $T=0$ relations $q=\qs$ and $\qs=2-\nu$ hold also for  
$0 \leq \frac{\mu}{J} < \frac{\chi}{2}$, whence in this interval one
simply
obtains
\begin{equation}
\nu = 2-\qs = 1,\quad \chi=\frac{2}{\sqrt{2 \pi q}}
\label{zeroTempeq6}
\end{equation}
\subsubsection{Density of states} 
\begin{figure}
\centerline{
\epsfxsize8cm
\epsfbox{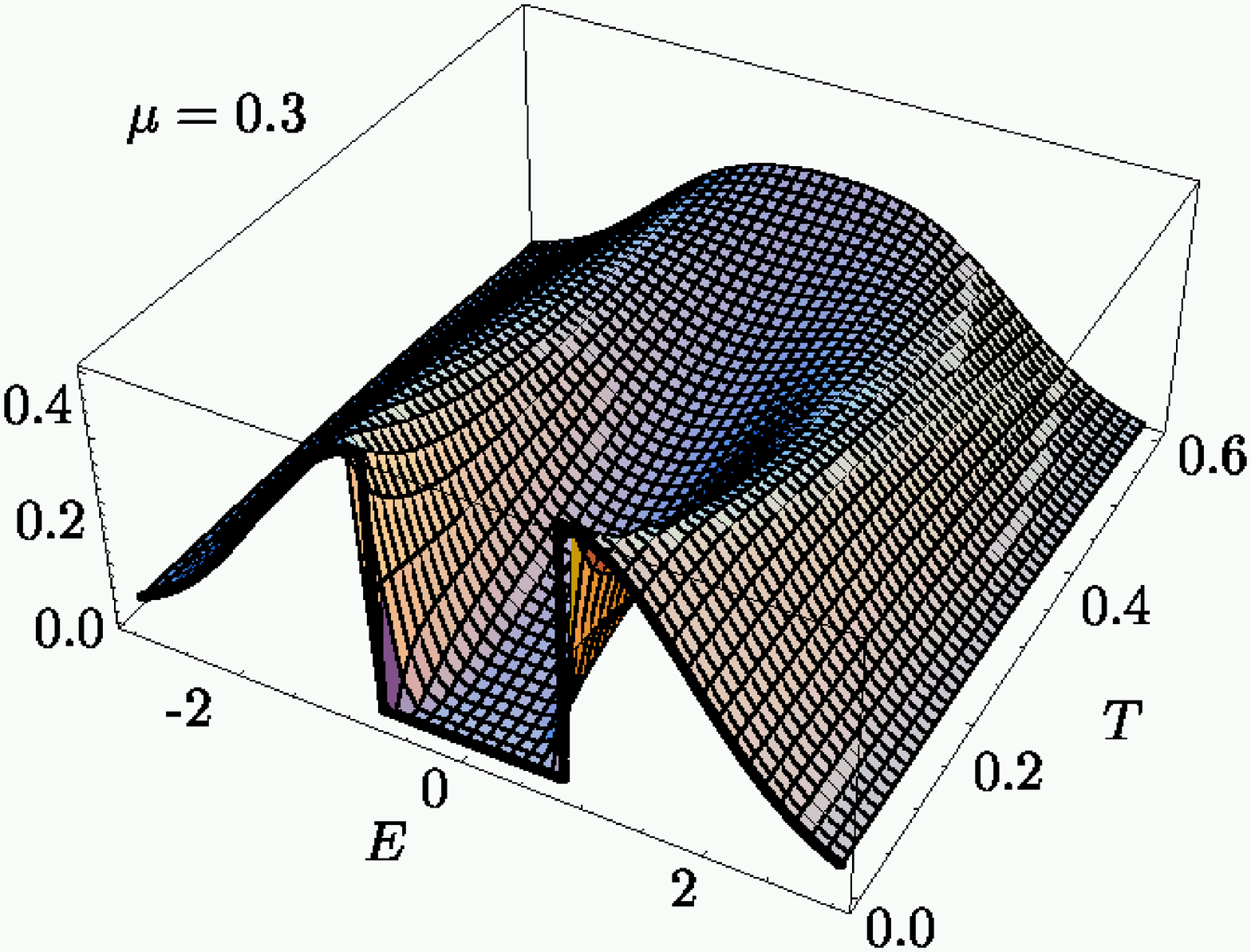}
\epsfxsize8cm
\epsfbox{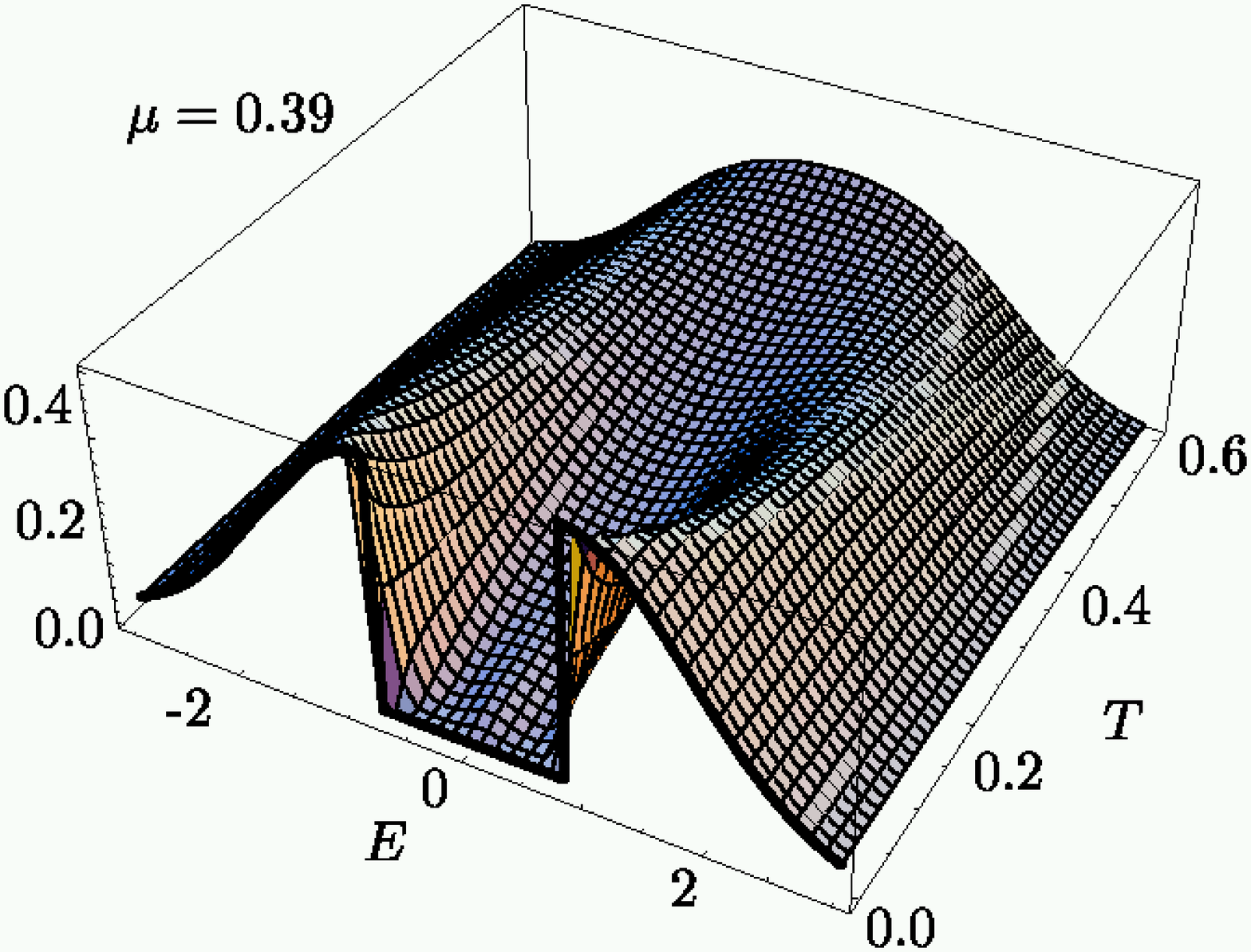}}
\centerline{
\epsfxsize8cm
\epsfbox{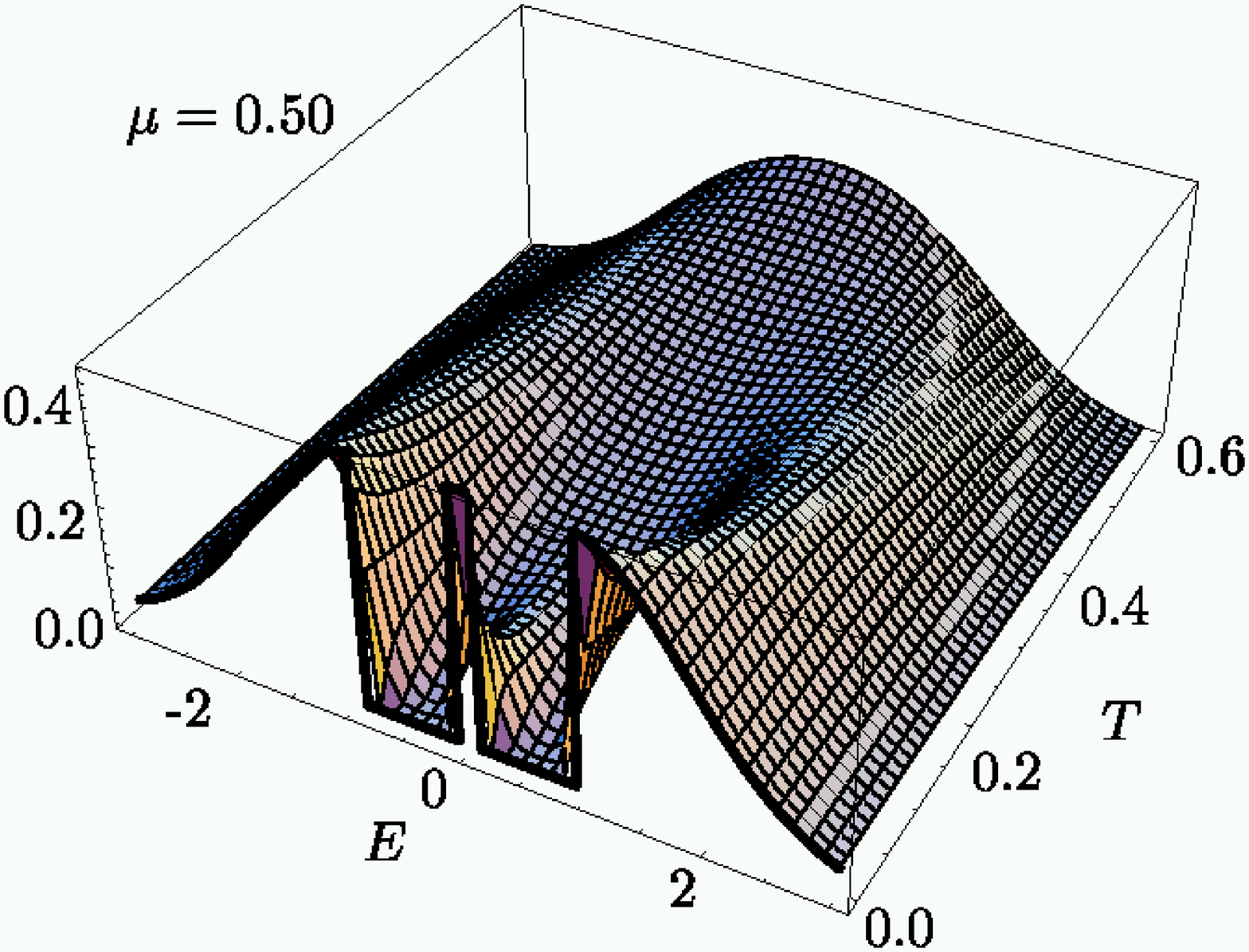}
\epsfxsize8cm
\epsfbox{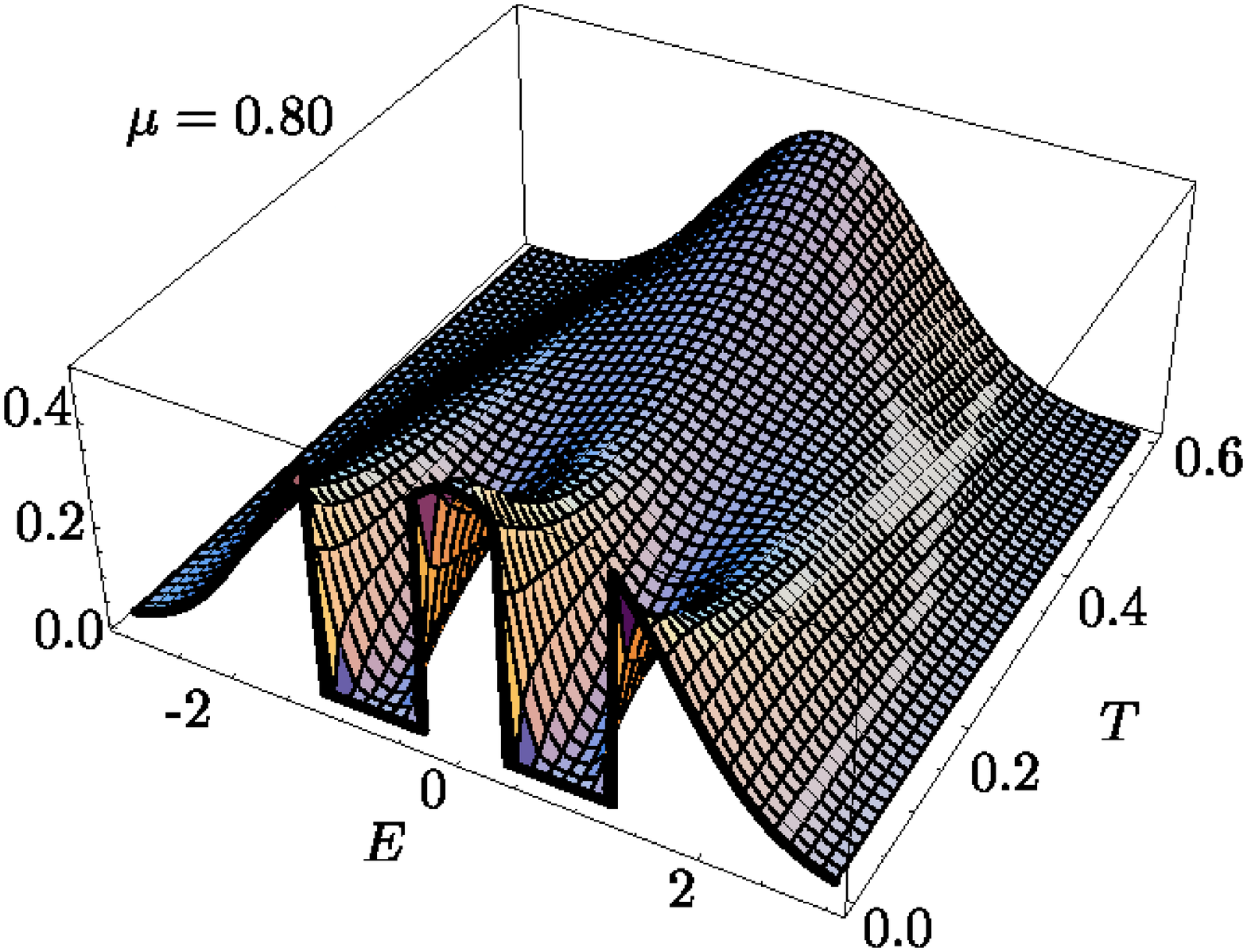}}
\caption{Density of states $\rho(E=\epsilon+\mu,T)$ as a function of
temperature below freezing,  
as calculated in the replica
symmetric approximation at specific chemical potentials $\mu$
(in units of $J$). The bold front line represents the independent $T=0$
result obtained from Equations (7-12).}
\label{0RSB.mu}
\end{figure}

The derivation of the Green's function from the generating functional
of spin glasses was discussed before \cite{oppermann98c}. 
The present work makes use of the same formalism, 
but evaluates it for all fillings. 
The (numerical) solutions for $q(\mu,T)$ and $\chi(\mu,T)$ are employed in
the calculation of the electronic density of states throughout the whole
spin glass phase.
The following set of Figures \ref{0RSB.mu} shows that a central band
emerges for high enough chemical potentials. 
In the 0RSB approximation the band gap discussed previously for half
filling \cite{oppermann98c} is visible up to $\mu =\frac{1}{2}
E_g^{(0)}=\frac{1}{\sqrt{2\pi}}$ and the DoS looks like one half of a
bath tub. 
When the chemical potential is moved towards the band gap value (we show
for example the DoS at $\mu = 0.39$), a tiny central band shows up at
low temperatures, but looses its weight again completely in the
$T\rightarrow0$ limit in favor of the magnetic side bands. 
Roughly speaking, a line given by $\partial^2 f/\partial\qs^2$=0, which,
starting at the tricritical point of the phase diagram, bends into the gap
edge \cite{oppermann99a} at $T=0$, wraps this small precursor of the central band. 
For chemical potentials exceeding the gap edge value
$\frac{1}{\sqrt{2\pi}}$, the central band manifests itself already at
higher $T$ and survives at $T=0$, becoming there a pure Gaussian
function of $E=\epsilon+\mu$ with finite height and symmetric cutoffs. 
In the 0RSB approximation both the central band and
the magnetic bands are sharply cut off and separated by gaps of identical
width $\chi$. As replica symmetry breaking will show, 
the gap size is given by the nonequilibrium susceptibility $\chib$, 
which agrees with $\chi$ only in this lowest order approximation - we
continue to discuss the gap in this chapter in terms of $\chi$. 
This quantity, which separates central from upper band and central from
lower magnetic band begins to vary with $\mu$ for
$|\mu| > \frac{1}{2}E_g^{(0)}$.
Figure \ref{0RSB.mu} combines the numerical finite
temperature calculations of 
$\rho(\epsilon,T|\mu)$ with the exactly calculated function
$\rho(\epsilon,T=0|\mu)$ given below, making use of the numerical
solutions for order parameter and susceptibility. As emphasized by the
thick lines at $T=0$, the $T=0$ density of states can be decomposed into 
isolated contributions of the three bands. Taking advantage of the
symmetry one may use the energy variable $E\equiv\epsilon+\mu$
\begin{equation}
\rho(T=0,E)= \rho_{-}(0,E) +
\rho_{c}(0,E) + \rho_{+}(0,E)
\end{equation}
where the central charge band is given by
\begin{equation}
\rho_c(0,E)=\frac{1}{\sqrt{2\pi
q(0,\mu)}}\hspace{.1cm}e^{-E^2/(2q(0,\mu))}\hspace{.15cm}
\Theta(\mu-\chi(0,\mu)/2-|E|)
\end{equation}
Upper and lower magnetic band contribute
\begin{equation}
\rho_{\pm}(E)=\frac{1}{\sqrt{2\pi q(0,\mu)}}
(e^{-(|E|-\chi(0,\mu))^2/(2q(0,\mu))}
\Theta(|E|-(\mu+\chi(0,\mu)/2)
\end{equation} 

\begin{figure}
\centerline{
\epsfxsize8cm
\epsfbox{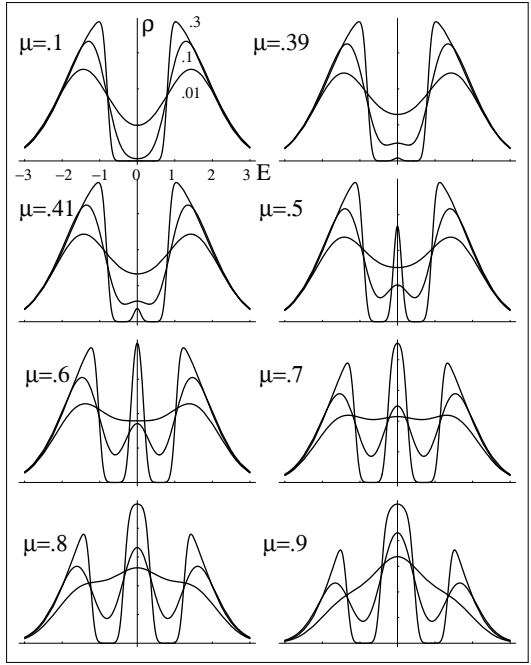}}
\vspace{.3cm}
\caption{The density of states shown for temperatures $T=.3, .1, .01$ at
chemical potentials $\mu=.1,.39,.41,.5,.6,.7,.8,.9$}
\label{collection}
\end{figure}

\begin{figure}
\centerline{
\epsfxsize8cm
\epsfbox{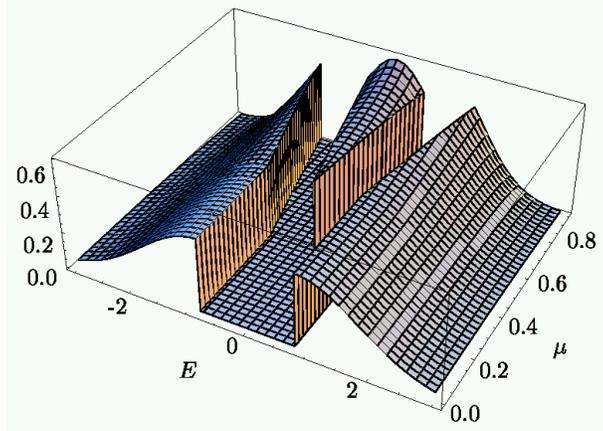}}
\caption{DoS at zero temperature as a function of $\mu$. 
Above $\mu=1/\sqrt{2 \pi}$, the central charge band appears. In this case, 
the Fermi level lies between the central and the upper magnetic band.}
\label{0RSB.T0}
\end{figure}

Let us first discuss the numerical results in the RS-approximation.
Figure \ref{collection} shows a collection at typical values
for the chemical potential:\\
i) Within the hard gap regime $0<|\mu|<1/\sqrt{2\pi}$, where
$\nu(T=0)=1$, a pronounced central band is absent; only in a small
range of low but finite temperatures a tiny midgap peak is observed. We
find that its existence is clearly linked to the characteristic line
mentioned above. This line separates the domain of the phase diagram,
where the free energy is minimized as a function of $\chi$, from the one
where it is maximized. This latter property is unrelated to the wellknown
maximization of $f$ by the SG order parameter $q$; in turn the presence of
$q$ is needed to render a solution with $\partial^2 f/\partial\chi^2<0$
stable.
For the present case we find a complex Almeida Thouless eigenvalue (for
replica-diagonal perturbation), which does at least not exclude stability
apart from the Parisi RSB.\\
ii) For chemical potentials large enough to sustain fillings different
from one electron per site, i.e. $|\mu|>\frac{1}{\sqrt{2\pi}}$, three bands
develop as the temperature falls below $T_f$ and become separated as
$T\rightarrow 0$. The Fermi level lies between the central charge band and
the upper magnetic band. The area under the central peak belongs to the
deviation from half-filling, described by $|\nu-1|$. The low $T$ results
(see Figure \ref{0RSB.T0})
confirm the numerical observation that the central band width is given by
$E_{cb}^{(0)} = 2\mu-\chi$ at $T=0$, that both left and right gap widths
obey $E_g^{(0)}=\chi$, while at half filling the relation reads
$E_g^{(0)}=2\chi$.\\
One may compare this approximate solution with an iterated perturbation
solution of the half--filled Hubbard model at zero temperature\cite{georges96a}: there a
hopping generated band shows up within the Hubbard gap insulating phase.\\

\subsection{Improved solutions with broken replica
symmetry and strong low--temperature effects}

\begin{figure}
\centerline{
\epsfxsize8cm
\epsfbox{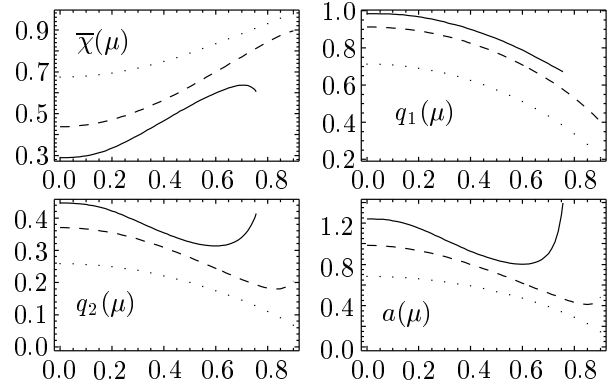}}
\vspace{.5cm}
\caption{Order parameters, nonequilibrium susceptibility, and Parisi
parameter, shown in 1RSB, as a function of $\mu$ at selected
temperatures $T=0.05$ (solid lines), $T=0.15$ (dashed lines),
and $T=0.3$ (dotted lines).} 
\label{figfiniteTparameters}
\end{figure}

One step replica symmetry breaking yields a large step towards
the exact solution: it allows to guess properties of the exact one quite
frequently.
As the basic starting formula for all thermodynamic properties we
rewrite the free energy in 1RSB in terms of parameters, which also allow
to obtain finite zero--temperature limits. It is therefore convenient to
use the nonequilibrium susceptibility, linear susceptibility, and Parisi
parameter denoted respectively by
$\chib=\beta J (\qs - q_1)$, $\chi=\chib+a (q_1-q_2)$, and $a=\beta
J m$ (we have chosen $\chib$, $\chi$, and $a$ as dimensionless).\\
Now the free energy density reads
\begin{align}
f=&\frac14 J (\chib(\frac{\chib}{\beta J} + 2 q_1 -2)+ a (q_1^2-q_2^2))
-T\ln 2 - \mu -\frac{J}{a}\int_{z_2}^G \ln \int_{z_1}^G 
\calC^{\frac{a}{\beta J}}\quad \mbox{with} \label{eq:f1RSBnewVariables}\\
\calC =& \cosh(\beta J (\sqrt{q_1-q_2}z_1 +\sqrt{q_2}z_2)) 
+\cosh(\beta\mu)e^{-\frac{\beta J}{2}\chib}
\end{align}
The condition of stationary free energy provides saddle--point equations
for the four parameters $\chib$, $q_1$, $q_2$, and $a$, which need to be
determined as functions of $\mu$ and $T$. The derivation of the thermal
selfconsistent equations is lengthy and hence omitted. Instead we present
in Figure \ref{figfiniteTparameters} the solutions as a function of $\mu$
for three selected values of temperature. For the use in the density of
states below, we have determined all necessary parameters almost
continuously on a grid of $(\Delta\mu/J,\Delta T/J)=(10^{-2},10^{-2})$.

The above formulation of the free energy allows to obtain the zero
temperature limit in terms of finite quantities
\begin{equation}
f=\frac12 J \chib (q_1 -1)+ \frac14 J a (q_1^2-q_2^2)
- \mu -\frac{J}{a}\int_{z_2}^G \ln \int_{z_1}^G \calC^{\frac{a}{\beta J}} 
\label{eq:f1RSB0T}  
\end{equation}
The zero temperature limit of the internal integral is calculated as
\begin{equation}
\begin{split}
\int_{z_1}^G \calC^{\frac{a}{\beta J}} \to&
\frac{1}{\sqrt{2\pi}} \int_{-\infty}^{\infty}dz_1 e^{-\frac{z_1^2}{2}}
e^{a |\sqrt{q_1-q_2}z_1 + \sqrt{q_2}z_2|} \\
=&
\frac12 e^{a \sqrt{q_2}z_2+ \frac12 a^2(q_1-q_2)}
\Bigl(1+\erf\bigl(\frac{\sqrt{q_2}z_2 + 
a (q_1-q_2)}{\sqrt{q_1-q_2}\sqrt{2}}\bigr)\Bigr)\\
&+\frac12 e^{-a \sqrt{q_2}z_2+ \frac12 a^2(q_1-q_2)}
\Bigl(1+\erf\bigl(\frac{-\sqrt{q_2}z_2 
+ a (q_1-q_2)}{\sqrt{q_1-q_2}\sqrt{2}}\bigr)\Bigr)
\end{split}
\label{eq:int1case1}
\end{equation}
for $\frac{\mu}{J}<\frac{\chib}{2}$ and
\begin{equation}
\begin{split}
\int_{z_1}^G \calC^{\frac{a}{\beta J}}
=&
\frac12 e^{a \sqrt{q_2}z_2 + \frac12 a^2(q_1-q_2)}
\Bigl(1+\erf\bigl(\frac{\sqrt{q_2}z_2 - (\frac{\mu}{J}-\frac{\chib}{2})
+a(q_1-q_2)}{\sqrt{q_1-q_2}\sqrt{2}}\bigr)\Bigr)\\
+&
\frac12 e^{a(\frac{\mu}{J}-\frac{\chib}{2})} 
\Bigl(\erf\bigl(\frac{(\frac{\mu}{J}-\frac{\chib}{2})
-\sqrt{q_2}z_2}{\sqrt{q_1-q_2}\sqrt{2}}\bigr)+\erf\bigl(\frac{(\frac{\mu}{J}
-\frac{\chib}{2})+\sqrt{q_2}z_2}{\sqrt{q_1-q_2}\sqrt{2}}\bigr) \Bigr)\\
+&
\frac12 e^{-a \sqrt{q_2}z_2 + \frac12 
a^2(q_1-q_2)}\Bigl(1+\erf\bigl(\frac{-\sqrt{q_2}z_2-(\frac{\mu}{J}
-\frac{\chib}{2})+a(q_1-q_2)}{\sqrt{q_1-q_2}\sqrt{2}}\bigr)\Bigr)
\end{split}
\label{eq:int1case2}
\end{equation}
for $\frac{\mu}{J}>\frac{\chib}{2}$.\\

\begin{figure}  
\centerline{ 
\epsfxsize8cm
\epsfbox{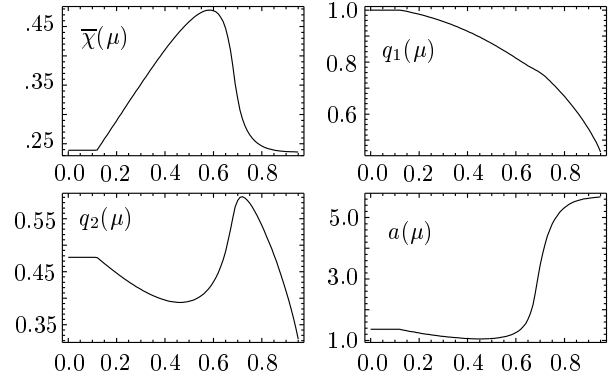}}
\vspace{.5cm}
\caption{Order parameters in 1RSB at zero temperature}
\label{fig_zeroTparameters}
\end{figure}

These preceding $T=0$--limits are the basic ingredients which we use
analytically to get the selfconsistent equations at zero temperature
in terms of finite parameters. Results of our final numerical
evaluation are collected in Figure \ref{fig_zeroTparameters}.
  
\begin{figure}
\centerline{
\epsfxsize8cm
\epsfbox{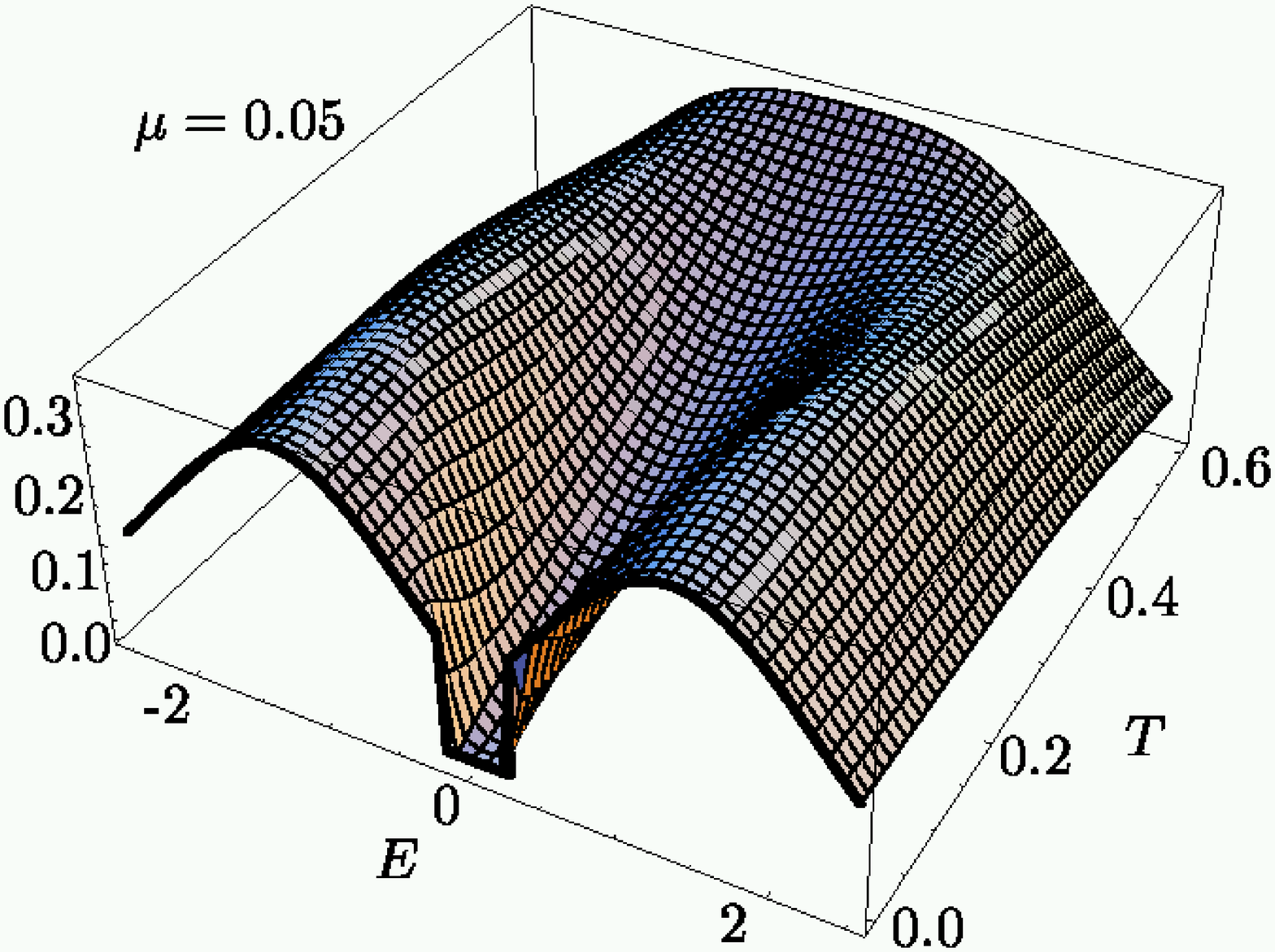}
\epsfxsize8cm
\epsfbox{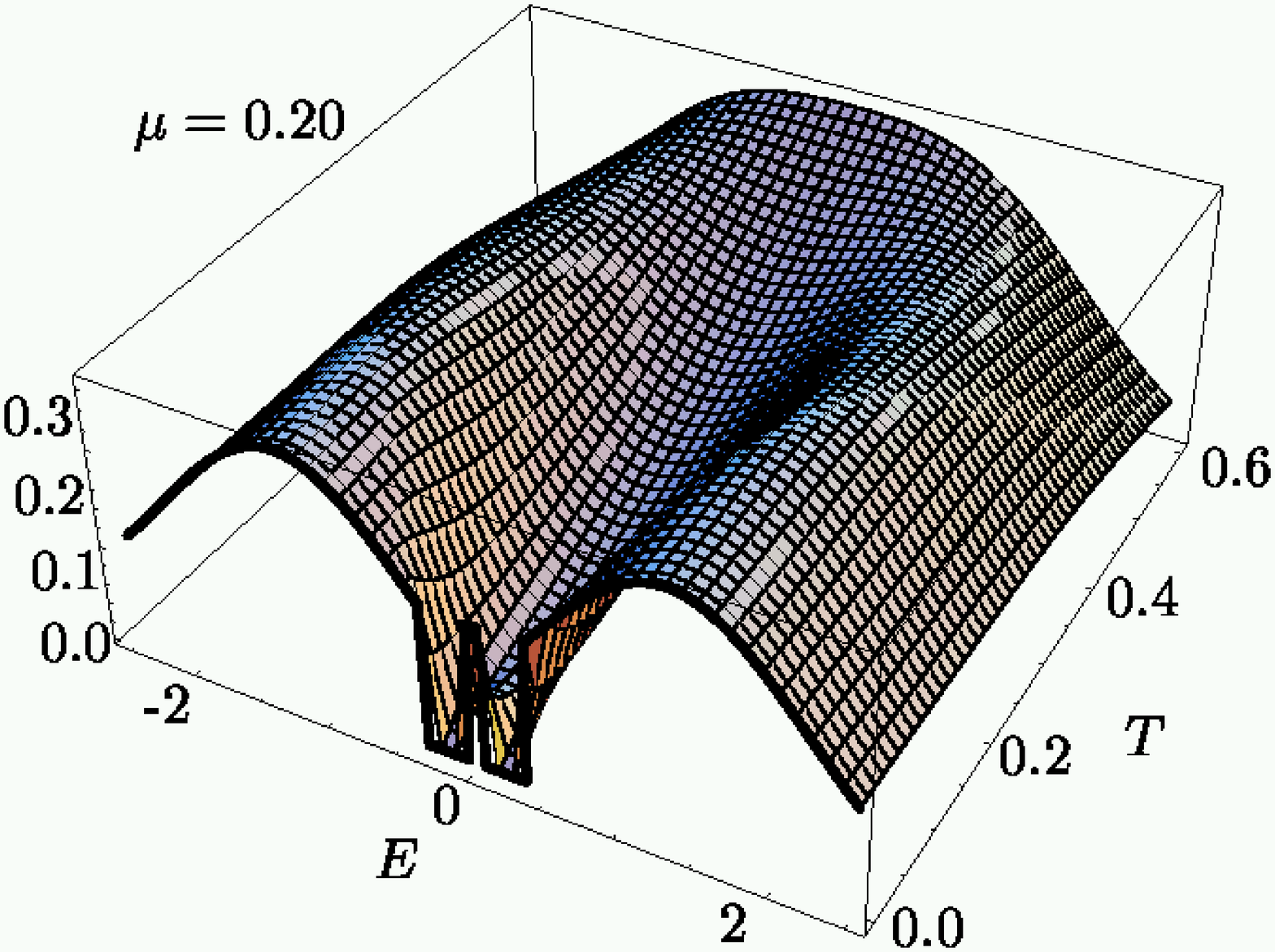}}
\centerline{
\epsfxsize8cm
\epsfbox{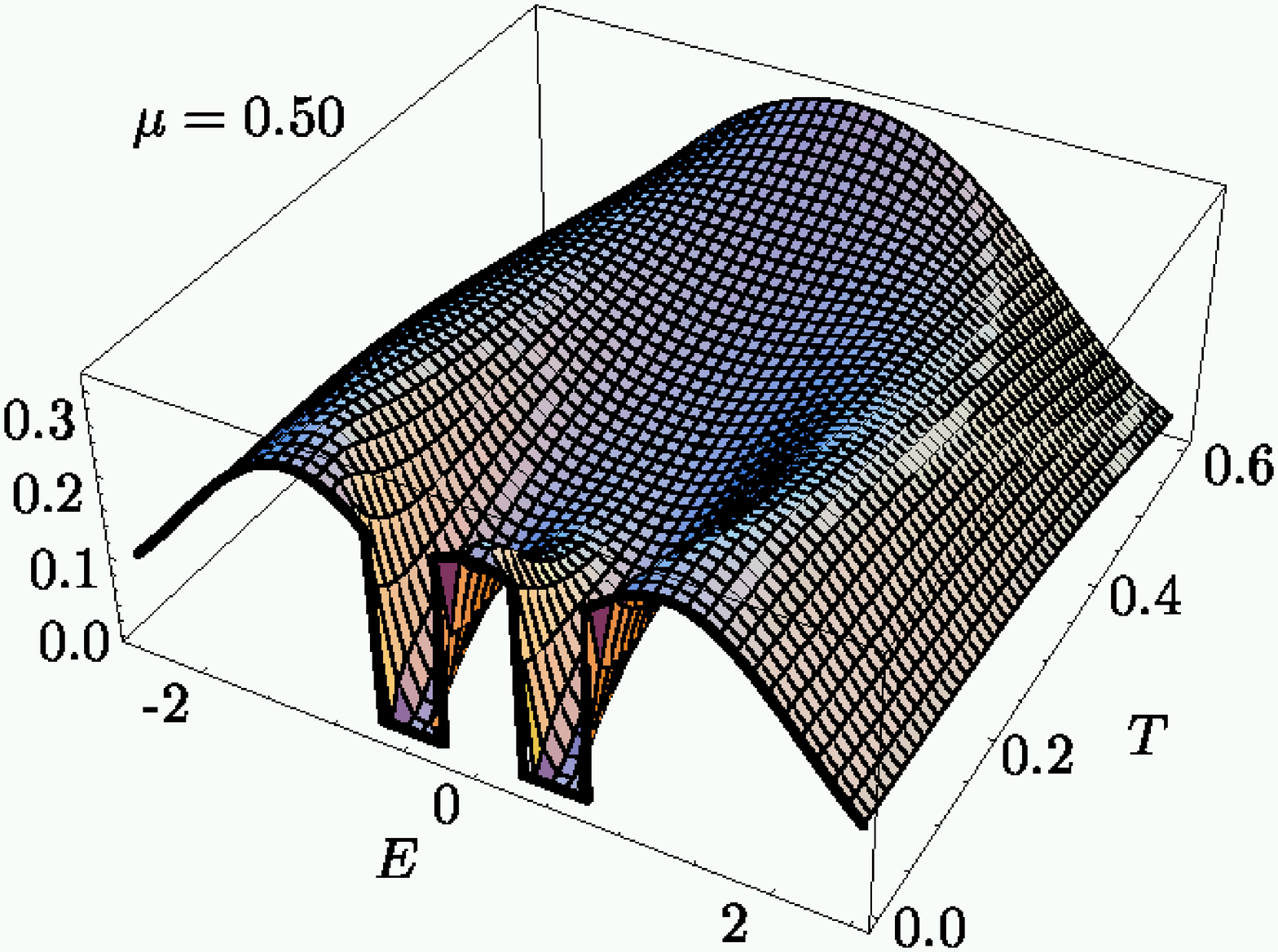}
\epsfxsize8cm
\epsfbox{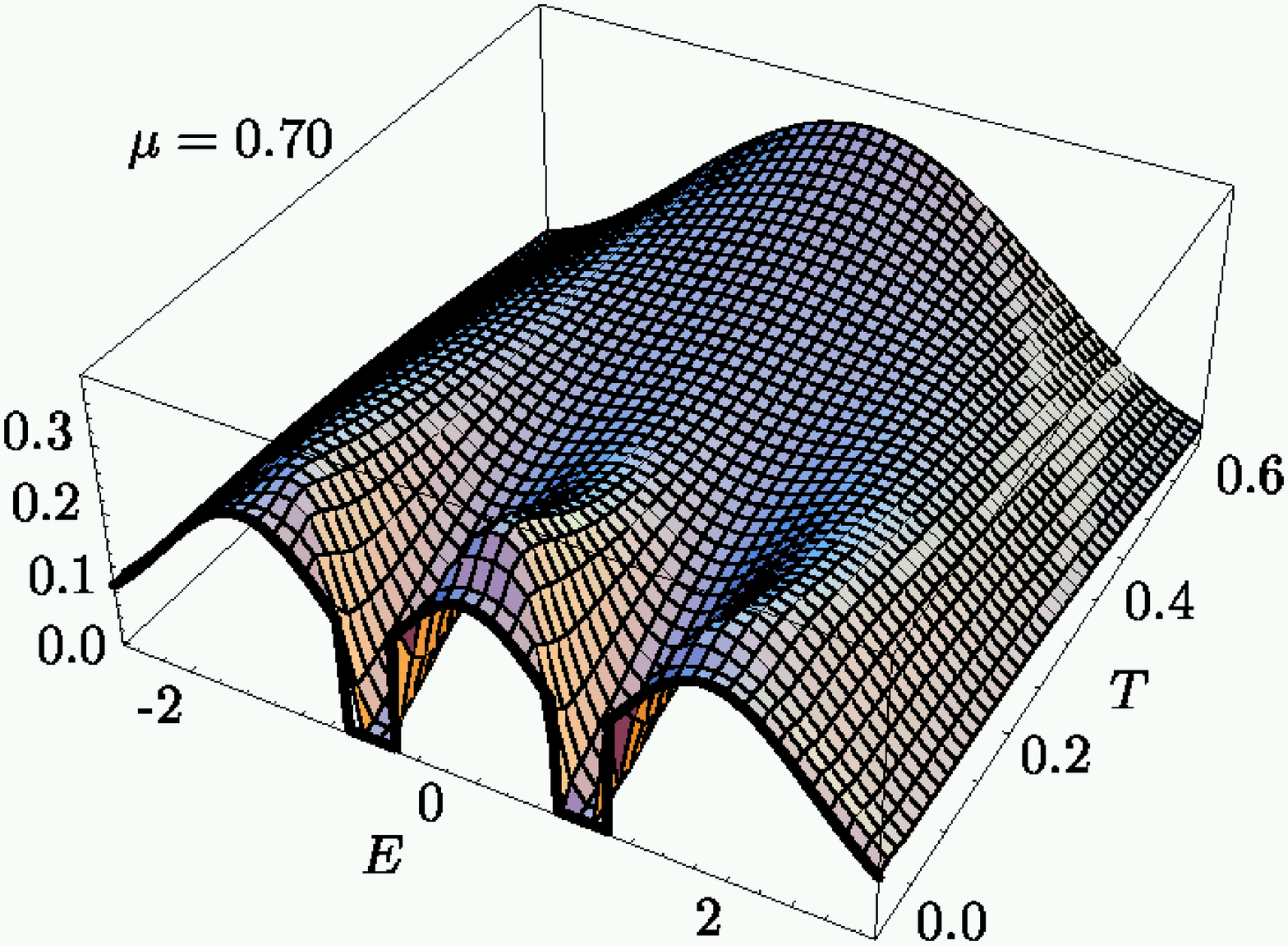}
}
\caption{Density of states at finite and at zero temperature in 1RSB shown
for $\mu=.05, .2, .5,$ and $.7$ as a function of energy $E=\epsilon+\mu$
and temperature $T$.}
\label{1RSB.mu}
\end{figure}

All of these parameters show a remarkable variation before the first order
transition regime to the paramagnetic state is approached at
$\mu_{c1}(T=0)\approx
0.881$. Even $q_1$,
which agrees with the spin autocorrelation function $\qs$ at $T=0$,
contains interesting behaviour. This will be extracted below in terms of
the change of the fermion filling under one--step replica symmetry
breaking. \\
The filling factor is an integrated quantity over the density of
states folded with the Fermi distribution. We can also exploit all
solutions, given so far for the order parameters, in order to determine
the fermionic density of states itself.
This is done first for all relevant chemical potentials and all
temperatures $O(10^{-2}) \leq T <T_f(\mu)$, resulting in the body of the
Figures \ref{1RSB.mu}. The $T=0$ solutions are then derived as follows.
Matching perfectly the results at lowest finite $T$, they are finally
combined with the body of each figure.\\

The density of states reads in 1RSB for $T>0$ and as a function of
$E\equiv \epsilon+\mu$
\begin{equation}
\rho(E) = 
\frac{1}{\sqrt{2 \pi}} \sqrt{\frac{\beta}{J \chib}}
(\cosh(\beta \mu) + \cosh(\beta E)) e^{-\frac{\beta J}{2}\chib}
\int_{z_2}^G \frac{\int_{z_1}^G \calC^{\frac{a}{\beta J}-1} 
e^{
-\beta (-E + (J \sqrt{q_2} z_2 + J \sqrt{q_1-q_2} z_1))^2/(2 J \chib)
}}
{\int_{z_1}^G \calC^{\frac{a}{\beta J}}}
\end{equation}

In the $T=0$ limit, the saddle point method allows to solve the internal
integrals exactly, which results in 
\begin{equation}
\rho(E) =
\frac{1}{\sqrt{2 \pi}}
\frac{1}{J \sqrt{q_1-q_2}}
\begin{cases}
  0 & \mbox{for}\; \mu < J \frac{\chib}{2} \; \mbox{and}\; 
  |E| < J \chib\\
  e^{\frac{a}{J}|E| - a \chib}  
\int_{z_2}^G\frac{e^{-\frac12 \frac{(|E| - J \chib - J
        \sqrt{q_2}z_2)^2}{J^2 (q_1-q_2)}}}{\int_{z_1}^G \calC^m} &
  \mbox{for}\;  \mu < J \frac{\chib}{2} \; \mbox{and}\; J \chib< |E|\\
  e^{a (\frac{\mu}{J} - \frac{\chib}{2})}\int_{z_2}^G\frac{e^{-\frac12 
  \frac{(J \sqrt{q_2}z_2 - E)^2}{J^2 (q_1-q_2)}}}{\int_{z_1}^G 
  \calC^m}& \mbox{for}\;  \mu > J \frac{\chib}{2} \; \mbox{and}\; 
  |E| < \mu - \frac{J \chib}{2}  \\
0 & \mbox{for}\;  \mu > J \frac{\chib}{2} \; \mbox{and}\;
\mu - \frac{J \chib}{2}  < |E| < \mu + \frac{J \chib}{2}\\
  e^{\frac{a}{J}|E|-a
      \chib}\int_{z_2}^G\frac{e^{-\frac12 \frac{(|E| - J \chib - J
        \sqrt{q_2}z_2)^2}{J^2 (q_1-q_2)}}}{\int_{z_1}^G \calC^m} & 
\mbox{for}\;  \mu > J \frac{\chib}{2} \; \mbox{and}\; \mu + \frac{J
    \chib}{2}< |E|
\end{cases}
\end{equation}
where Eqs. (\ref{eq:int1case1}) and (\ref{eq:int1case2}) have to be substituted.

\begin{figure}
\centerline{
\epsfxsize8cm
\epsfbox{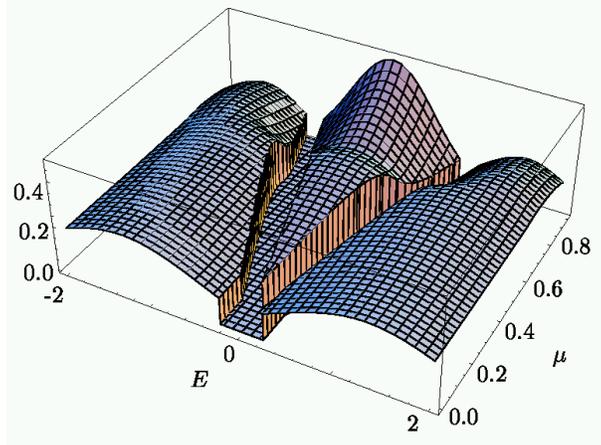}}
\caption{DoS at zero temperature for 1RSB. 
The central band appears above $\mu \approx 0.119$.}
\label{1RSB.T0}
\end{figure}

Figure \ref{1RSB.mu} displays the exact selfconsistent evaluation 
at one step replica symmetry breaking.
The analytical intermediate solutions described before, allowed to
include the $T=0$ solution into the Figure. The full picture shows the
band structure at all temperatures below $T_f$. The choice of $\mu$ values
covers a wide range from almost hal filling at $\mu=0.05$ to $\mu=0.7$
through the entire spin glass phase.\\
A comparison with the lowest order approximation, shown in Figure
\ref{0RSB.mu}, combined with analytical results for $\infty$RSB, 
gives a clear hint for the exact $ISG_f$ solution.\\
Only the first part at $\mu=0.05 < \frac{1}{2}E_g^{(1)}\approx .119$ 
does not contain the central band, since $\nu(\mu=.05,T=0)=1$ in 1RSB. 
For $E_g^{(1)}=.119 < \mu < E_g^{(0)}=\sqrt{\frac{2}{\pi}}$ the central
band is present
in 1RSB, while it was absent in 0RSB in this interval. This effect sets in
at low temperatures, corresponding to the smaller magnetic energy scales
set by the new additional order parameter of 1RSB.
Above temperatures determined roughly by the random field condition
$\partial^2 f/\partial {q^{aa}}^2 = 0$ ($\tilde{q}\equiv q^{aa}$), which
corresponds to a random field crossover line $T_{rf}(\mu)$, the spectral
weight does not show a peak at $E=0$. 
It needs larger values of $\mu$ to find a large central peak
already for intermediate temperatures below $T_f$. On the other hand,
the central band is already well developed at $\mu=0.5$ for low enough
temperatures. At this value of $\mu$ the lowest order approximation leads
only to a very small band, since $\mu=0.5$ exceeds only by little the
0RSB gap edge.\\
As for half filling the ratio between the gap widths and the (finite) DoS
at the gap edge is constant. For higher order symmetry breaking the gap
shrinks and the spectral weight at the edge diminishes correspondingly;
thus spectral weight is moved into parts of the gap region, as the
approximation is improved step by step.\\
The crossover from finite low $T$ to the exact $T=0$ solutions, shown by
fat lines in the 3D-plots, redistributes considerably the spectral weight.
At higher $\mu$, for example $\mu=0.7$ as shown in Figure \ref{1RSB.mu},
the central band shows a maximum also as a function of temperature. The
DOS height in the center decreases again at lowest temperatures, the band
becomes broader at $T=0$.\\
Comparing the $T=0$ limit of Figures 1 and 2, which show the 0RSB
approximation, with that of Figure \ref{1RSB.mu} for the first improved
1RSB approximation, one finds that replica symmetry breaking leads to a
refinement of the band structure. Each further RSB step, until the exact
$\infty$RSB solution is reached, will modify magnetic and nonmagnetic
bands according to the still smaller scales and magnetic order parameters.
However, although one step RSB is not yet exact in this model, further
refinements are much smaller in size and it is possible to imagine the
exact result from our Figure \ref{1RSB.mu}.\\  
As a function of a continuously varying chemical potential, the 1RSB
solution for the zero temperature density of states is displayed in Figure
\ref{1RSB.T0}.
The central band maintains the wedge-like shape already observed in
the 0RSB result. Its thick end however shows a new structure near the
discontinuous magnetic breakdown, while the other end progressed by a
large step towards $\mu=0$.
The distance from $\mu=0$ will be further reduced in kRSB with $k>1$.  
\subsubsection{Infinite breaking of replica symmetry}
The top of the wedge reaches $\mu=0$ in $\infty$RSB and at the same time
the gap widths approach zero; the DoS becomes zero at $\pm\mu$ but
stays finite yet very small near $E=\pm\mu$. 
The derivation is analogous to the one presented in Ref.
\onlinecite{oppermann98c}:
The ratio between gap width DoS value at the gap edge is invariant under
RSB and hence they decay together to zero in the limit of infinite RSB.
Here it is assumed that the nonequilibrium susceptibility, which
determines the gap width at any order $k$ of the RSB, approaches zero
for $k\rightarrow\infty$ as in the half-filled case. The latter result was
inferred from the work of Thouless, Anderson, and Palmer \cite{thouless77b}.

\section{The connection between fermion concentration and replica symmetry
breaking}
\begin{figure}
\centerline{  
\epsfxsize8cm
\epsfbox{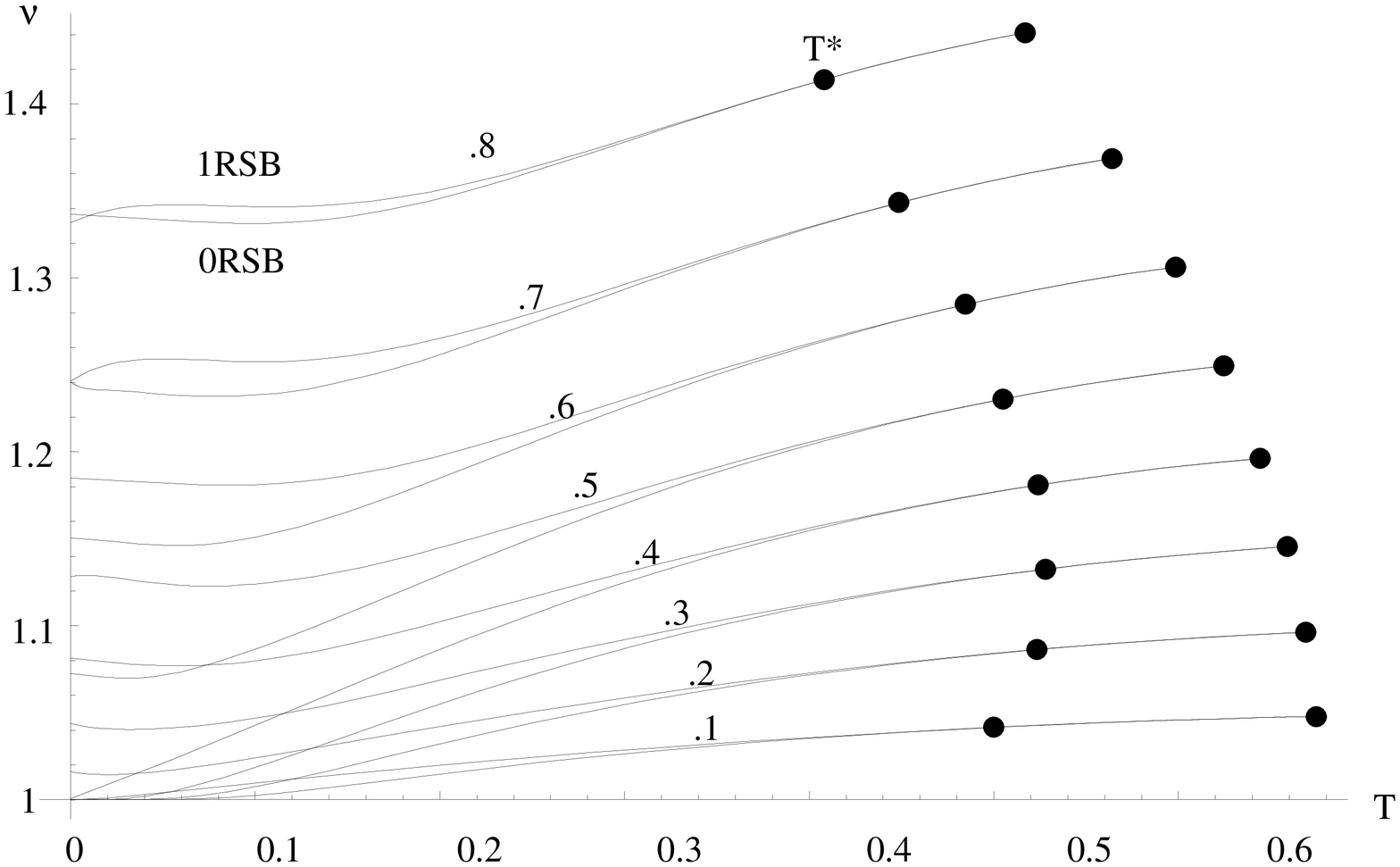}}
\vspace{1cm}
\caption{Results for the temperature dependence of the filling factors in
0RSB and 1RSB are grouped together for $.1\leq\mu\leq .8$ with
$\Delta\mu=.1$. The right endpoints are the freezing temperatures
$T_f(\mu)$, while at the intermediate points, denoted by $T^*$ on the pair
of lines for $\mu=.8$, the fillings $\nu_{0RSB}$ and $\nu_{1RSB}$ differ
by only $10^{-4}$. Zeros of $\nu_{0RSB}-\nu_{1RSB}$ exist within
$T^*(\mu)<T<T_f(\mu)$ and at low $T$ for $\mu>.7$} 
\label{fillingT}
\end{figure}

The filling factor can be described by the summation over the
imaginary frequency Green's function ${\cal{G}}$ by
\begin{equation}
\nu=T\sum_{\epsilon_n,\sigma}{\cal{G}}_{\sigma}(\epsilon_n)e^{i\epsilon
0_+}
\end{equation}
The Green's function is related to the density of states discussed before
by the usual spectral representation 
${\cal{G}}(\epsilon_l)=\int
d\epsilon\frac{\rho(\epsilon)}{i\epsilon_l-\epsilon}$ (and by
$\rho(\epsilon)=-\frac{1}{\pi}Im[G^R(\epsilon)]$), which means that the
whole fermion propagator changes under RSB. Still the summation over all
frequencies could either wipe out or maintain
this dependence. Indeed what we find is
a transition between these two alternatives inside the ordered phase.\\ 
We have emphasized the role of replica symmetry breaking for quantum
dynamics and low energy excitations. Despite the absence of quantum
dynamics in charge correlations, and the absence of spin-charge couplings
in the Hamiltonian, we find that replica symmetry breaking affects
spin-correlations and charge-average in a qualitatively different way.
This is remarkable also because $\nu$ is a global quantity (summed over
all frequencies) in contrast to the Green's function (which is local in
$\epsilon$).\\ 
In one-step breaking we report in this section a crossover line, which
separates a regime of almost invisible RSB-effects in the filling
factor and in $\tilde{q}$ from one with large RSB-effects in these
quantities. The magnetic order parameter does not show any sign of this
crossover. The announcement of these effects evokes on one hand the old,
resolved problem \cite{usadel90a} of absence of RSB below the
freezing temperature, and on the other hand the phenomenon of a
Gabay-Toulouse line \cite{toulouse81a}, followed by a crossover to a region
with RSB effects in all order parameters \cite{sherrington82a}.\\ 
The first case bears no relation with our
case: the present model is static in charge and spin-correlation and the
particular role of dynamic effects \cite{usadel90a}, which occured in
the transverse field Ising model, do not exist here.\\   
The crossover line describing the onset of RSB in transversal
correlations of a Heisenberg spin glass in a
magnetic field however, has a vague resemblence, provided we imagine
charge degrees of freedom as transversal with respect to spin. The
chemical potential then roughly corresponds to the magnetic field in the
standard case. However a detailed mapping between the two models does not seem
feasible.\\ 
In Figure \ref{fillingT} the fermion filling factor is shown.
Pairs of ($\nu_{0rsb(\mu)}$,$\nu_{1rsb}(\mu)$) are grouped together for
$\mu=.1,.2,...,.8$.
The detailed plots 
contain two interesting features: 
each pair of lines seems to merge asymptotically, but the lines still
cross each other at a $T^*_0(\mu)$, staying close together for
$T^*_0(\mu)<T<T_f(\mu)$.
Since the lines cross, having almost identical slope, it is difficult to
determine with sufficient numerical precision the line $T^*_0(\mu)$ of
crossings. 
We therefore chose points $T^*(\mu)$ where 0RSB-- and 1RSB--lines differ
only by $10^{-4}$. In between these points and the endpoints at the
freezing temperature corresponding to the $\mu$-parameter of each curve, 
the 0RSB and 1RSB curves cross at least once. \\ 
The interesting fact now is that there is a region below the freezing
temperature, where replica symmetry breaking effects in the fermion
filling factor almost vanish. While this occurs in the charge related
quantity, the magnetic observables such as order parameters still show
large RSB effects. One may find it more surprising that RSB
effects appear at all in the fermion concentration, but it seems still more
important that these effects disappear almost (or perhaps completely in
the exact solution of infinite RSB) within the spin glass phase.\\ 
We do not attempt here to discuss any result for itinerant models, but it
is nevertheless clear that the feature described for the insulating model
can have implications on itinerant systems.\\
In addition to the separation of magnetic and nonmagnetic bands
described before for the ordered phase, one also finds a different
exposure of charge- and spin-related quantitites to the effects of replica
symmetry breaking. None of the quantities is excluded from this, despite
the fact that the model does not contain a spin-charge interaction. The
only coupling is mediated by the chemical potential.\\  

\begin{figure}
\centerline{
\epsfxsize8.5cm
\epsfbox{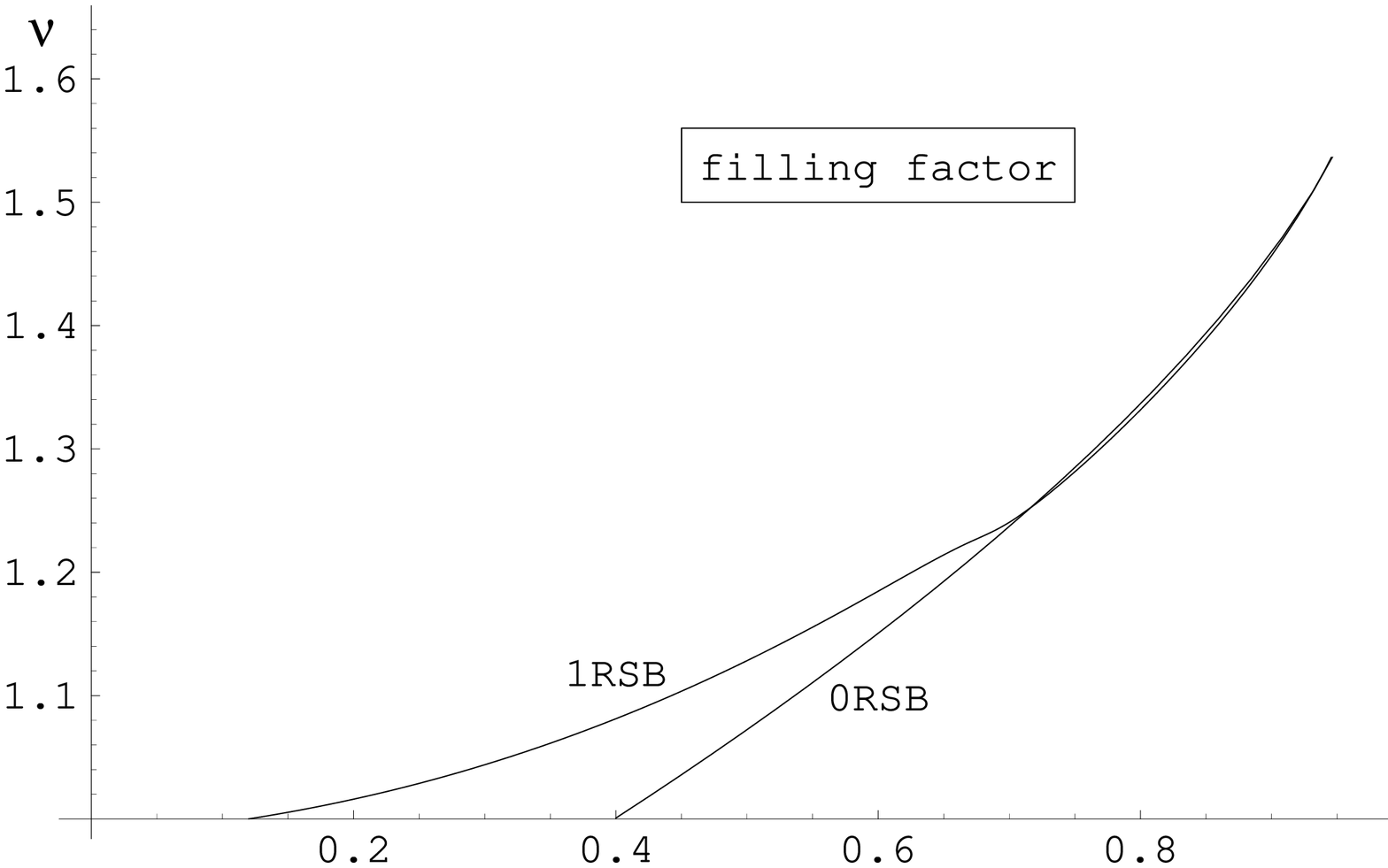}
\epsfxsize10cm
\epsfbox{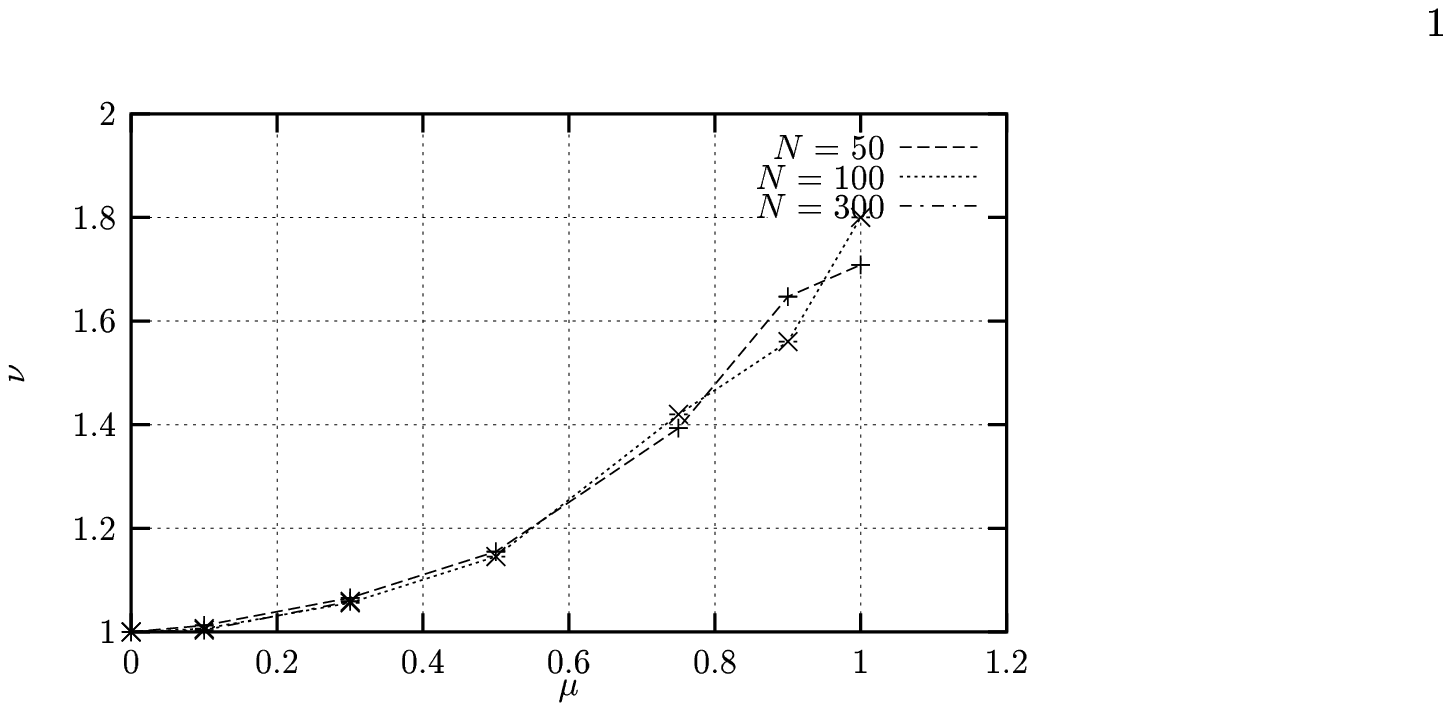}
}
\vspace{.5cm}
\caption{The filling factor at $T=0$ are shown as a function of the
chemical potential $\mu$. The 0RSB- and 1RSB-results of this paper are
included in the figure on the left hand side, while on
the right the numerical solution of the corresponding TAP--equations is
taken from Ref.\cite{rehker99a} for comparison. There, $N$ denotes the
system size.} 
\label{zeroTfilling}
\end{figure}

Beyond the crossing point of the two approximations at $\mu^*\approx0.7J$ both
lines stay close together as if no symmetry breaking effect would occur at
all between $\mu^*$ and the discontinuous breakdown of magnetic order at
$\mu_1\approx0.881$. On the other side, for $\mu<\mu^*$ the RSB effect is
large. Moreover, the right hand side of Fig.\ref{zeroTfilling} shows for
comparison the numerical result for the corresponding TAP-equation of
finite size systems \cite{rehker99a}. 
This generalized TAP--result should correspond to the full Parisi solution
with $\infty$RSB. At $T=0$ the filling $\nu$ starts to differ from one,
when the chemical potential moves through the gap edge. The gap size
depends on the number of RSB steps and decreases to zero in the
$\infty$RSB solution. Thus it is clear that left endpoint of the
filling curve $\nu(\mu)$ moves into $(\nu=1,\mu=0)$. The change from the
calculated 1--step RSB solution to this exact one is smaller than the one
from 0RSB to 1RSB. The shape of the 1RSB solution resembles almost
perfectly the one found numerically from the generalized TAP-equation.
Also quantitative agreement is obtained for $\mu$ smaller than roughly
$0.5$. The nonvanishing kRSB--corrections for $k=2$ and higher can be
expected to be almost invisibly small. 
Deviations in the high-$\mu$ region might be due to numerical
problems of the algorithm in \onlinecite{rehker99a} either because of the vicinity of the first order phase
transition or due to finite size effects.\\
In general however the TAP-solution is already in good agreement with the
analytical 1RSB solution, whose extension to the full RSB solution is
obvious. The flat increase of $\nu(\mu)$ from $1$ (probably with slope
zero) in the generalized Parisi solution is consistent with the fact that
for rare nonmagnetic regions the central band does not start as a
$\delta$-peak, but as function with finite height and a width increasing
smoothly from zero as $\mu$ becomes finite.  \\ 

\section{Summary and outlook}
By analyzing the spin glass phase in the $(\mu,T)$-plane we found
two magnetic and one central nonmagnetic band, which become separated
perfectly at $T=0$. Their separation by finite gaps in any finite RSB
approximation turns into a pseudogap separation in the exact
infinite-step RSB solution, assuming that the nonequilibrium
susceptibility vanishes as predicted by the TAP-solution. \\
RSB effects did appear in the magnetic
order parameters, in the density of states, and in the quantum dynamics
displayed by the Green's functions, and also in the fermion filling
(being at $T=0$ simply the integrated DoS).\\
The results presented here for the $ISG_f$ model trigger speculations not
only on other insulating random interaction models, like XY-- and
Heisenberg quantum spin glasses, but also on itinerant extensions.\\
For example, the photoconductivity of the itinerant extension of the
 model is related to
the integrated overlap of frequency--shifted density of states. The
filling factor proved that integration does not wash out RSB--effects,
which could therefore be expected to exist in the photoconductivity at low
temperatures.
In itinerant systems the possibility of localization in the pseudogap
regime of small DoS will be decisive for ac-- and dc--conductivity.
Again RSB-effects will emerge and mark the theory of localization due to a
frustrated random magnetic interaction in the spin glass phase. It will be
highly interesting to compare this with the fully frustrated Hubbard model
in infinite dimensions 
\cite{kotliar99a}.\\ 
While any finite step RSB does not yet allow to state that the exact line
$T^*(\mu)$ must be different from $T_f(\mu)$ (as it is in 1RSB) and that
RSB perhaps exactly vanishes in the region $T^*(\mu)<T<T_f(\mu)$, this
remains a serious option.\\
For itinerant models the question arises whether an RSB transition,
disconnected from the magnetic transition, can also occur in metallic spin
glasses and perhaps will influence transport properties.

\section{Acknowledgments}
We thank E. P. Nakhmedov for discussions.
This work was supported by the Deutsche Forschungsgemeinschaft under
contract Op28/5-1 and by the SFB410. One of us (H.F.) also acknowledges
support by the Villigst foundation.


\begin{thebibliography}{10}

\bibitem{parisi80b}
G. Parisi, J. Phys. A {\bf 13},  1887  (1980).

\bibitem{parisi80c}
G. Parisi, J. Phys. A {\bf 13},  L115  (1980).

\bibitem{binder86a}
K. Binder and A. Young, Rev. Mod. Phys {\bf 58},  801  (1986).

\bibitem{fischer91a}
K.~H. Fischer and J. Hertz, {\em Spin Glasses} (Cambridge University Press,
  Cambridge, 1991).

\bibitem{georges96a}
A. Georges, G. Kotliar, W. Krauth, and M. Rozenberg, Rev. Mod. Phys. {\bf 68},
  13  (1996).

\bibitem{oppermann98c}
R. Oppermann and B. Rosenow, Europhys. Lett. {\bf 41},  525  (1998).

\bibitem{oppermann96a}
B. Rosenow and R. Oppermann, Phys. Rev. Lett. {\bf 77},  1608  (1996).

\bibitem{oppermann99a}
H. Feldmann and R. Oppermann, Eur. Phys. J. B {\bf 10},  429  (1999).

\bibitem{oppermann99b}
H. Feldmann and R. Oppermann, preprint  (1999).

\bibitem{daCosta94a}
F.~A. da~Costa, C.~S.~O. Yokoi, and S.~R.~A. Salinas, J. Phys. A {\bf 27},
  3365  (1994).

\bibitem{thouless77b}
D. Thouless, P. Anderson, and R. Palmer, Phil. Mag. {\bf 35},  593  (1977).

\bibitem{usadel90a}
G. B\"uttner and K.~D. Usadel, Phys. Rev B {\bf 41},  428  (1990).

\bibitem{toulouse81a}
M. Gabay and G. Toulouse, Phys. Rev. Lett. {\bf 47},  201  (1981).

\bibitem{sherrington82a}
D.~M. Cragg, D. Sherrington, and M. Gabay, Phys. Rev. Lett. {\bf 49},  158
  (1982).

\bibitem{rehker99a}
M. Rehker and R. Oppermann, J. Phys.: Condens. Matter {\bf 11},  1537  (1999).

\bibitem{kotliar99a}
G. Kotliar, cond-mat/9903188  .

\end{thebibliography}
\end{document}